# The Spherical Bolometric Albedo of Planet Mercury

Anthony Mallama

14012 Lancaster Lane

Bowie, MD, 20715, USA

anthony.mallama@gmail.com

2017 March 7

Abstract

Published reflectance data covering several different wavelength intervals has been combined and analyzed in order to determine the spherical bolometric albedo of Mercury. The resulting value of 0.088 +/- 0.003 spans wavelengths from 0 to 4 μm which includes over 99% of the solar flux. This bolometric result is greater than the value determined between 0.43 and 1.01 μm by Domingue et al. (2011, Planet. Space Sci., 59, 1853-1872). The difference is due to higher reflectivity at wavelengths beyond 1.01 μm. The average effective blackbody temperature of Mercury corresponding to the newly determined albedo is 436.3 K. This temperature takes into account the eccentricity of the planet's orbit (Méndez and Rivera-Valetín. 2017. ApJL, 837, L1).

1. Introduction

Reflected sunlight is an important aspect of planetary surface studies and it can be quantified in several ways. Mayorga et al. (2016) give a comprehensive set of definitions which are briefly summarized here. The geometric albedo represents sunlight reflected straight back in the direction from which it came. This geometry is referred to as zero phase angle or opposition. The phase curve is the amount of sunlight reflected as a function of the phase angle. The phase angle is defined as the angle between the Sun and the sensor as measured at the planet. The spherical albedo is the ratio of sunlight reflected in all directions to that which is incident on the body. The geometric and spherical albedos are functions of wavelength. Finally, the Bond albedo, as it is called by Mayorga et al., is the integral of the spherical albedo over all wavelengths and, thus, it represents the bolometric value.

In this paper the term *spherical bolometric albedo* is used instead of Bond albedo because the latter term is employed inconsistently in the published literature. In some cases the Bond albedo refers to a wavelength-specific quantity while in others it is integrated over wavelength. To be specific, the spherical bolometric albedo as used here is the fraction of incident sunlight that is reflected in all directions and at all wavelengths. As such, it enters into the energy balance equation for the planet.

The study of reflected sunlight from Mercury dates back to the $19^{th}$ century. These early efforts were improved upon with photoelectric photometry in the mid-$20^{th}$ century. Data acquired during the $21^{st}$ century goes further still and includes measurements made from spacecraft and from ground-based CCD instruments. Domingue et al. (2010) summarized this long history of observations and listed all the pertinent references, so that discussion will not be repeated here.

In a subsequent paper Domingue et al. (2011) determined an albedo (referred to in that paper as 'bolometric') of 0.081 for Mercury over the wavelength range of 0.43 to 1.01 µm which includes 59% of the solar flux. The purpose of this paper is to calculate a more complete spherical bolometric albedo over more than 99% of the solar flux by analyzing the planet's reflected light from 0 to 4 µm.

This paper examines and reviews various wavelength regions to ultimately derive the spherical bolometric albedo over 99% of the solar flux and to determine Mercury's effective temperature. Section 2 reviews the visible and far-red albedo determined by Domingue et al. (2011) spanning 0.43 to 1.01 µm. Section 3 addresses the UV and blue wavelengths from 0.00 to 0.42 µm. Section 4 considers the near-IR from 1.02 to 1.46 µm. Section 5 describes the thermal-IR range from 1.47 to 4.00 µm. Section 6 gives the spherical bolometric albedo derived over all of the above wavelength ranges and provides an estimate of its uncertainty. Section 7 discusses the uncertainties, lists the planet's effective blackbody temperature and briefly describes Mercury's spectrum where reflection and emission overlap. Section 8 provides a summary and discusses the conclusions.

2. Albedo from 0.43 to 1.01 µm

Domingue et al. (2011) evaluated data from the Mercury Dual Imaging System (MDIS) on the MESSENGER spacecraft. Disk-integrated data was obtained over phase angles that ranged from 53 to 145 degrees. Those authors computed phase integrals and geometric albedos at the eleven MDIS wavelengths. The product of these two quantities gives the spherical albedo, using the terminology in this paper, although it is called the Bond albedo in their paper. To be clear, the quantities they reported are the ratios of light reflected in all directions to incident light at each of the 11 MDIS wavelengths.

Next, Domingue et al. (2011) convolved the solar irradiance spectrum with each MDIS filter and fit that as a function of wavelength with a sixth-order polynomial. The spherical albedos were also fit to a sixth-order polynomial. The bolometric albedo was then taken to be the average spherical albedo weighted by the spectral irradiance of the Sun. This resulted in a value of 0.081 over the wavelength range of 0.43 to 1.01 µm.

3. Albedo from 0.00 to 0.42 µm

In order to evaluate the spherical albedo at wavelengths below the MDIS spectral range described in the previous section, data from the Mercury Atmospheric and Surface Composition Spectrometer (MASCS) on the MESSENGER spacecraft as reported by Domingue et al. (2010) was used in this study. In particular the numerical spectral data plotted in their figure 8 for three observed phase angles was kindly provided by D. L. Domingue (private communication). Data from the MASCS instrument starts at 0.306 µm, so the albedo at shorter wavelengths had to be determined by extrapolation. This section begins by discussing the observed data from 0.31 to 0.42 µm and then describes the extrapolation below 0.31 µm.

The data plotted in figure 8 of Domingue et al. (2010) are 'reflectance' or I/F quantities. These values were converted to fluxes in watts per square meter of collecting area per meter of wavelength (W/m^3) using the following equation from Domingue et al. (2010)

$$I/F = ( R / R_{sun} ) * ( \pi / \Omega )$$

Eq. 1

where $R$ is the flux from Mercury, $R_{sun}$ is the flux of the Sun on the planet and $\Omega$ is the solid angle of Mercury at a distance of 1 AU. The upper panel of Fig. 1 shows the Mercury I/F values for all three of the phase angles reported as well as the solar flux. The lower panel shows Mercury's flux as indicated by $R$ in Eq. 1. Computations were performed at intervals of 0.01 µm for this analysis and for all the other similar analyses in this paper.

The first step in the computation of spherical albedo is determination of the geometric albedo. Observations of Mercury made at smaller phase angles are preferred for this calculation because the planet's phase curve is very steep. Thus the ratio of the observed to the geometric quantity grows rapidly with phase angle and so does the conversion factor. According to the phase curve determined by Mallama et al. (2002) the factors for the three phase angles reported by Domingue et al. (2010) are 8.70 for phase angle 74.0 degrees, 12.99 for 87.2 degrees, and 14.49 for 91.0 degrees. So, the factor of 8.70 was applied to the fluxes at phase angle 74.0. The results were then multiplied by the phase integral determined by Mallama et al. (2002), 0.478, to give the flux from Mercury over the entire 4π steradians of the sphere. Next, the set of spherical values was integrated over the spectral range 0.31 to 0.42 µm which resulted in a flux of 6.0 W/m^2 for Mercury. (This again takes into account the planet's solid angle at a distance of 1 AU.) The solar flux integrated over the same wavelength range was found to be 139.2 W/m^2. Finally, the integrated Mercury flux was divided by the integrated solar flux resulting in an albedo of 0.043 which is the weighted average from 0.31 to 0.42 µm. The values of Mercury flux, solar flux and albedo quoted above, as was well as the percentage of solar flux in the stated wavelength integral (10.2%), are shown in the second line of data in Table 1.

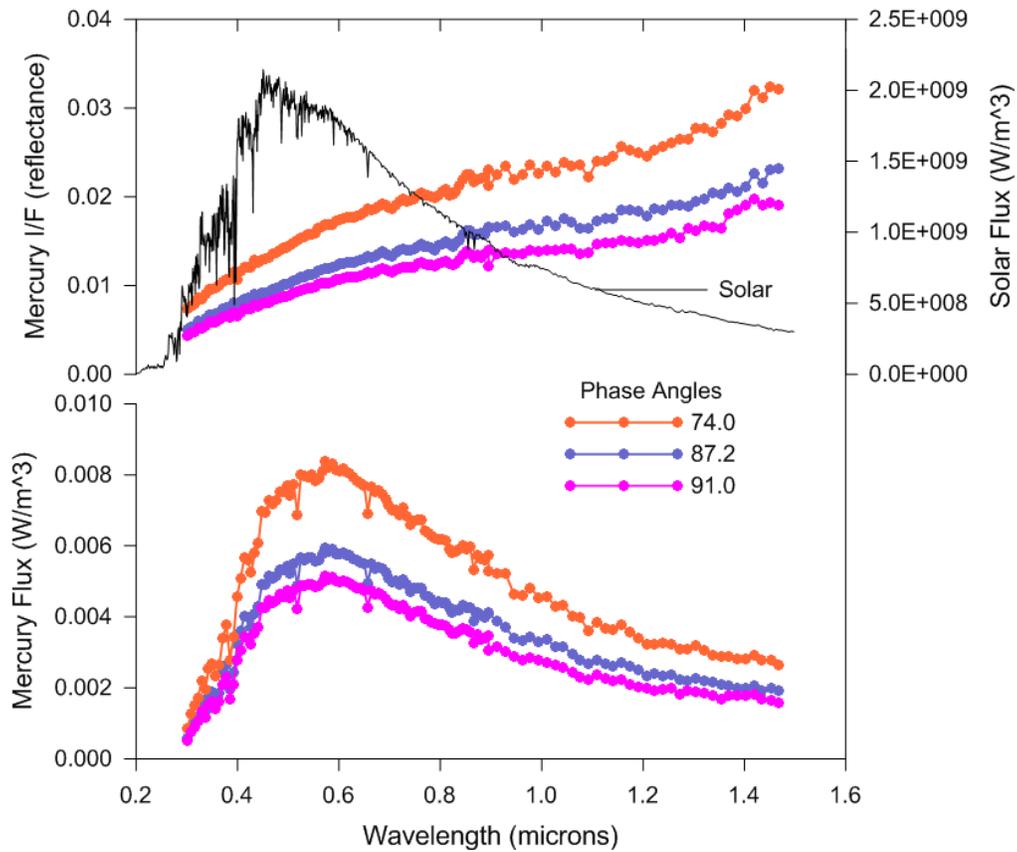

Fig. 1. Top panel: The reflectance of Mercury steadily increases with wavelength while the flux of the Sun reaches a peak near 0.45 µm. Bottom panel: The reflected flux of Mercury peaks near 0.57 µm due to the decreasing flux of the Sun. Mercury reflectance value are MASCS data derived by Domingue et al. (2010). Solar fluxes are from Wehrli (1985).

Table 1. Mercury Spherical Albedos (by Wavelength) and the Spherical Bolometric Albedo

| Wavelength (um) | Data Source | Albedo Source | Solar % | W/m^2 Mercury | W/m^2 Sun | Albedo |
|---|---|---|---|---|---|---|
| 0.00 - 0.30 | Domingue et al. 2010 * | This paper | 1.2 | 0.2 | 17.0 | 0.011 |
| 0.31 - 0.42 | Domingue et al. 2010 | This paper | 10.2 | 6.0 | 139.2 | 0.043 |
| 0.43 - 1.01 | Domingue et al. 2011 | Domingue et al. 2011 | 58.9 | 65.2 | 804.7 | 0.081 |
| 1.02 - 1.46 | Domingue et al. 2010 | This paper | 16.0 | 23.0 | 218.6 | 0.105 |
| 1.47 - 2.45 | McCord and Clark 1979 | This paper | 10.4 | 18.9 | 142.7 | 0.132 |
| 2.46 - 4.00 | McCord and Clark 1979 * | This paper | 2.6 | 6.5 | 34.9 | 0.186 |
| | * Extrapolated | | | | | |
| Bolometric: | | | 99.3 | 119.6 | 1357.0 | 0.088 |

Having examined the fluxes and albedos from 0.31 to 0.42 µm directly from MASCS data, it was then necessary to extrapolate Mercury fluxes below 0.31 µm. Fig. 2 illustrates that the extrapolated flux is extremely small. The first line of data in Table 1 indicates Mercury's flux as 0.2 W/m^2 while the flux of the Sun is 17.0 W/m^2 (just 1.2% of its total). So the weighted albedo over this spectral range is only 0.011.

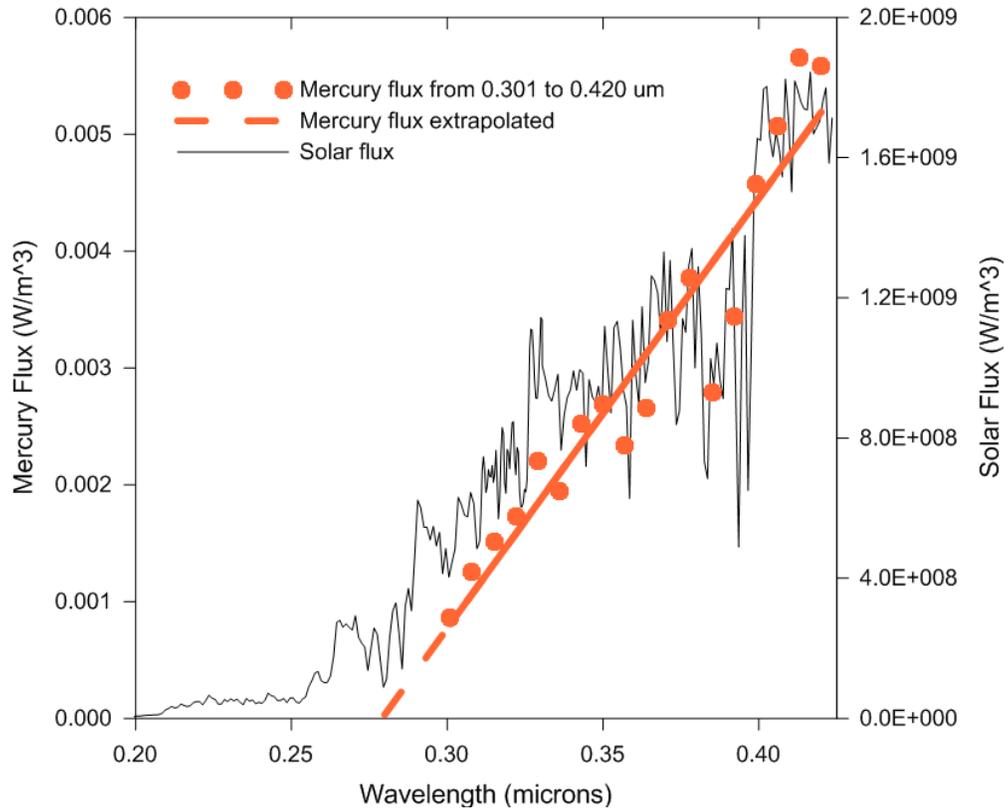

*Fig. 2. The observed flux of both Mercury and the Sun diminish very sharply toward shorter wavelengths in the UV portion of the spectrum. The extrapolated flux of Mercury below 0.301 µm is exceedingly small. Mercury fluxes are from this study based on MASCS reflectance data derived by Domingue et al. (2010). Solar fluxes are from Wehrli (1985).*

4. Albedo from 1.02 to 1.46 µm

The same method described in Section 3 for analyzing short-wavelength data was applied to wavelengths from 1.02 to 1.46 µm. Briefly, the MASCS I/F values for phase angle 74.0 derived by Domingue et al. (2010) were converted to fluxes as indicated in Fig. 1. Those observed flux values were, in turn, converted to spherical values. However, there is a complication at these longer wavelengths because the high surface temperature on the day-side of Mercury contributes measurable thermal flux beyond one micron. Therefore, this thermal contribution must be removed from the observed flux in order to isolate the reflected component associated with the planet's albedo.

Planck's law was used to compute the thermal contribution to the observed flux values. The projected area of the day-side of Mercury was taken to be the fraction of its disk illuminated at phase angle 74.0 degrees for this computation. Fig. 3 shows the thermal flux for a range of effective temperatures from 550 to 650 K in relation to the total observed flux. The thermal flux for the highest temperature, 650 K, would account for all of the observed flux at 1.46 µm leaving no budget for reflected flux. Therefore the actual effective temperature must be lower.

In order to determine the actual effective temperature, the observations of Mercury recorded by McCord and Clark (1979) were evaluated. These data span wavelengths from 0.65 to 2.50 µm and were reported with and without thermal flux as described in the next section. For the purposes of the analysis in this section, however, the dataset of McCord and Clark was matched to the MASCS dataset from 0.65 to 1.00 where thermal flux is negligible. (This 'matching' is described in the next section.) Then the excess of the MASCS fluxes relative to the McCord and Clark reflection-only fluxes in the wavelengths above 1.0 µm was attributed to thermal flux in the MASCS observations. The magnitude of this excess indicated an effective temperature of 575 K.

After removal of thermal flux using Planck's law and an effective temperature of 575K, the integrated reflected flux of Mercury from 1.02 to 1.46 µm based on the MASCS data was determined to be 23.0 W/m^2. The flux of the Sun in this wavelength span is 218.6 W/m^2 (16.0% of its total). The corresponding weighted albedo is 0.105. These values are listed in the fourth line of data in Table 1.

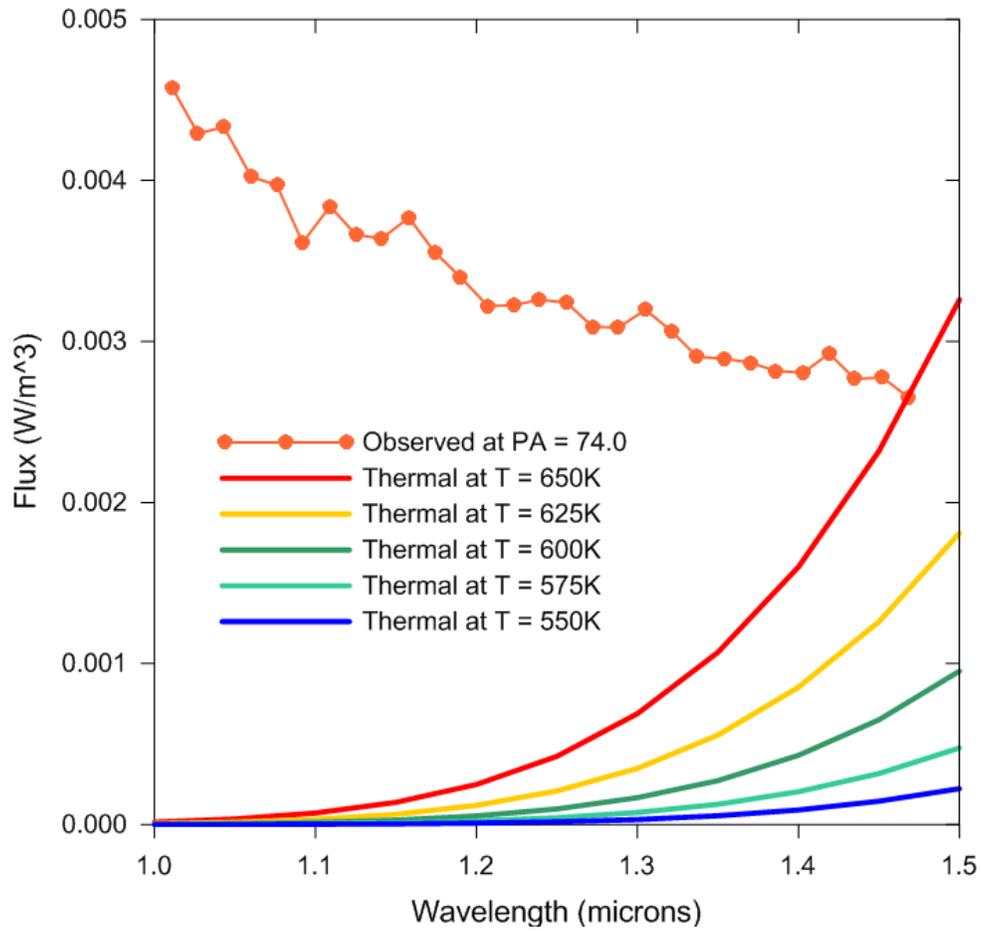

*Fig. 3. Thermal flux values derived from Planck's law for surface temperatures from 550 to 650 K are compared to the observed fluxes derived from the MASCS data reported by Domingue et al. (2010). All data correspond to phase angle 74.0 degrees.*

5. Albedo from 1.47 to 4.00 µm

McCord and Clark (1979) obtained long wavelength spectra of Mercury from Mauna Kea, Hawaii on three nights. Determination of Mercury's reflectance employed calibration observations of the star β Gem and took into account the ratio of intensity between that star and the Sun. Data reduction included the removal of significant atmospheric extinction and the authors considered that process to be most successful for the spectrum obtained on 1976 April 21. Mercury also happened to be at phase angle 77.4 degrees on that date which is conveniently close to the 74.0 degree phase angle data from the MASCS spectrum analyzed in this paper.

The spectra recorded by McCord and Clark extended from 0.65 to 2.50 µm. The flux is almost exclusively thermal at the long end of these wavelengths which allowed those authors to model that thermal component and to remove it from the entire spectrum. The remaining (reflected) amount is plotted in their figure 2 and that graph was digitized for the present analysis. A question arises immediately, though, because the values of reflectance are much higher than those from MASCS. The average ratio between the two from 0.65 to 1.00 µm is 6.97 and there is no significant variation with wavelength. While the root cause for this variation could not be determined, application of the ratio cited above gave a very good fit between the two data sets. The top panel of Fig. 4 shows the correspondence between reflectance values after applying the ratio but before correcting the MASCS data for the thermal component. (Correction of the MASCS data for thermal flux was discussed in Section 4.) The bottom panel illustrates the excellent agreement between the flux values after correction of MASCS for thermal flux.

The data of McCord and Clark extends to 2.50 µm. However, that beyond 2.45 µm appears to be very noisy so it was not used in this analysis. Likewise, the region of overlap with MASCS data from 0.65 to 1.46 was also omitted from the analysis in this section because it has already been taken into account in the albedo derived from MASCS data in Section 4.

The flux values of McCord and Clark from 1.47 to 2.45 µm were analyzed in the same way as were the MASCS data as described in the sections above. The reflected flux of Mercury was found to be 18.9 W/m^2, while that of the Sun in this same interval is 142.7 W/m^2 (10.4% of its total). Thus the weighted average albedo from 1.47 to 2.45 µm is 0.132, as indicated in the fifth line of data in Table 1.

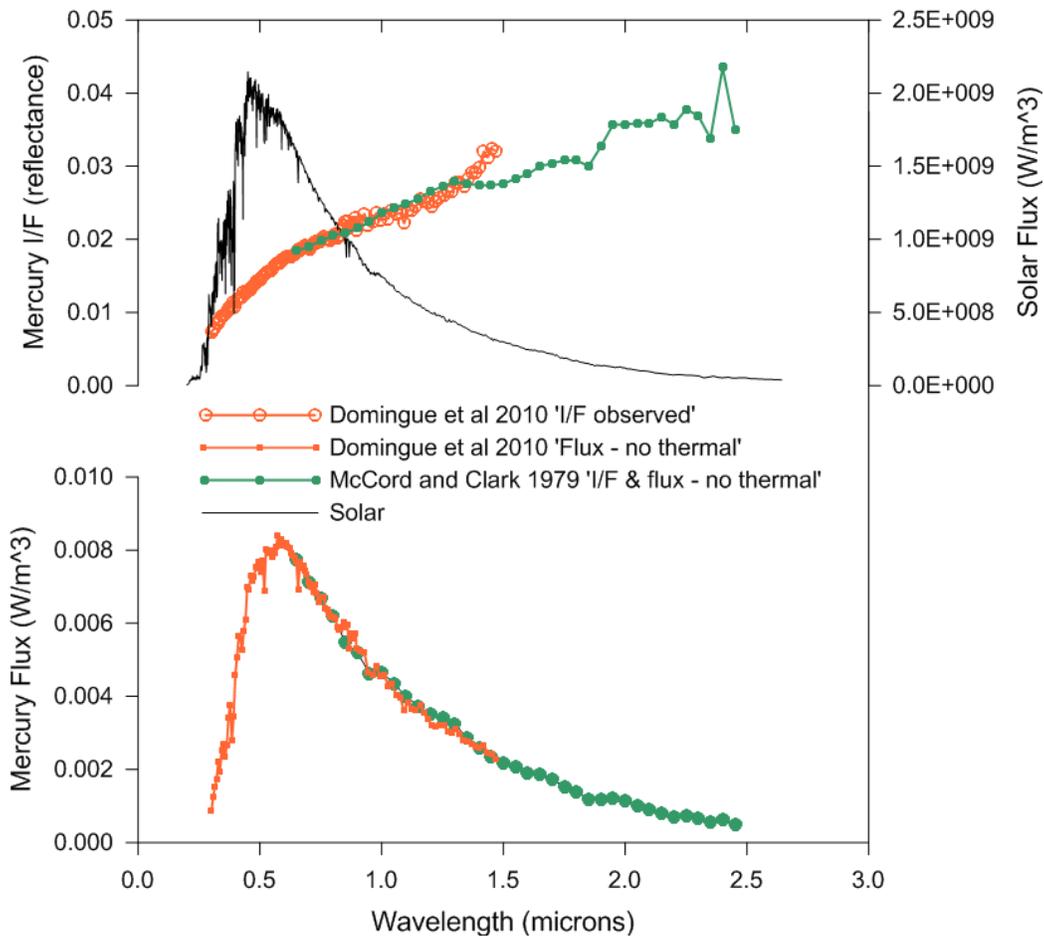

*Fig. 4. Top panel: The correspondence between reflectance values after dividing the McCord and Clark (1979) values by 6.97 but before correcting the MASCS data reported by Domingue et al. (2010) for the thermal component. The excess in the MASCS data at about 1.5 μm is due to this thermal component. Bottom panel: Excellent agreement between the flux values of both data sets at all overlapping wavelengths after removing the thermal flux (using Planck's law) from the MASCS data.*

The Sun emits 96.7% of its flux below 2.45 μm, leaving 3.3% unaccounted for in the analysis described up to this point. So, reflectance was extrapolated to 4.00 μm from the data between 1.25 and 2.45 μm of McCord and Clark as shown in Fig. 5. That extrapolation now accounts for 99.3% of the solar flux. The reflected flux of Mercury in the extrapolated interval from 2.46 to 4.00 was found to be 6.5 W/m^2, while that of the Sun in this same interval is 34.9 W/m^2 (2.6% of its total). Thus the weighted average albedo from 2.46 to 4.00 μm is 0.186, as indicated in the sixth line of data in Table 1.

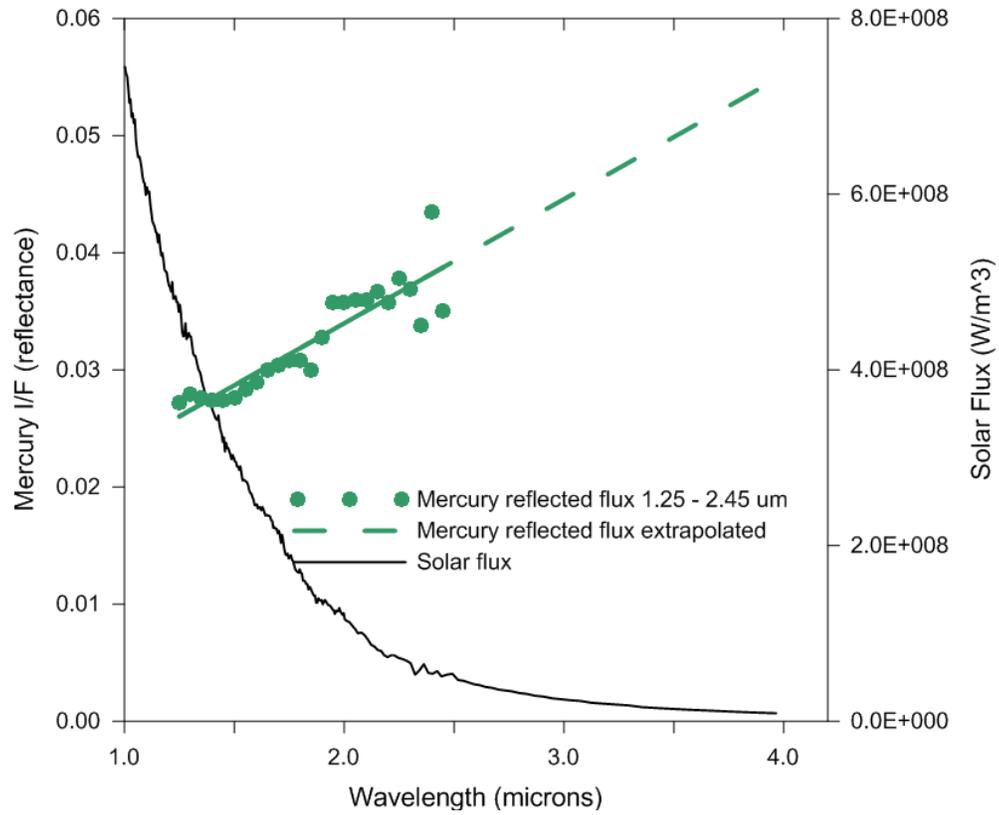

*Fig. 5. Circles indicate the observed reflectance as derived in this paper from the data reported by McCord and Clark (1979). The solid line is the best linear fit to those observations while the dashed line is the extrapolation out to 4.0 µm. The solar flux (Wehrli, 1985) is seen to be sharply declining.*

6. Spherical bolometric albedo and uncertainty

The last line of Table 1 indicates that the sum of the non-thermal fluxes from Mercury from 0 to 4 µm is 119.8 W/m^2, while that from the Sun is 1357.0 W/m^2. The corresponding weighted mean albedo is 0.088. Since the solar flux over this range of wavelengths is 99.3% of its total over all wavelengths the value 0.088 may be taken to be a very close approximation to the spherical bolometric albedo of Mercury.

The uncertainty is estimated by wavelength interval as follows. In the regions where extrapolation was used (0.00 - 0.30 and 1.47 - 2.45 µm) the error is likely to be the largest and a value of 20% is assigned. In the spans where data from Domingue et al. (2010) and from McCord and Rogers (1979) were used without extrapolation (0.31 - 0.42, 1.02 - 2.45 µm) the error is estimated to be a smaller 10%. In the region where Domingue et al. (2011) very carefully determined the albedo (0.42 - 1.01) an error of 5% is assigned. The rationale for assigning these particular uncertainties is discussed in the next section.

These percentage uncertainties were multiplied by the flux values for Mercury listed in Table 1. The root-sum-square (RSS) of those products was found to be 4.6 W/m^2 which is 3.8% of the total 119.8 W/m^2 flux of Mercury. When this percentage is applied to the spherical bolometric albedo of 0.088, the uncertainty of the albedo is found to be +/- 0.003.

7. Albedo uncertainties and effective temperature

This discussion addresses the uncertainties in the derived albedos in more detail and computes an effective blackbody temperature for Mercury based on the albedo. In addition the general characteristics of Mercury's spectrum are discussed.

The uncertainties cited in Section 6 were estimated as follows. A new spherical albedo of 0.075 was derived for the wavelength region from 0.42 to 1.01 µm from MASCS data using the method described in this paper. That value was then compared to the albedo of 0.081 reported by Domingue et al. (2011) from MDIS data which is 8% higher. So, a generous uncertainty of 10% was assigned to the results derived in this paper for the wavelength intervals 0.31 to 0.42 and 1.02 to 2.45 µm. In fact, the method used here, which depends on visible light phase information, should give reasonably accurate results across a wide range of wavelengths since Mercury has no atmosphere to impart significant color effects as a function of phase angle. One indication of this insensitivity to wavelength is that the phase integral listed by Domingue et al. (2011) for the shortest and longest MDIS wavelengths (0.43 and 1.01 µm) are very similar at 0.443 and 0.446, respectively. For the wavelength regions were extrapolation was used, specifically, 0.00 to 0.30 and 2.46 to 4.00 µm , the 10% uncertainty was doubled to 20% to provide an added margin for the possible error due to extrapolation itself.  Domingue et al. (2011) does not provide an estimate for the uncertainties in their spherical albedo. However, they explicitly determined phase integrals, geometric albedos, and spherical albedos for each of the 11 MDIS filters which leaves very little room for error. So, an uncertainty of 5% was assigned to their value for the wavelength region from 0.42 to 1.01 µm.

No uncertainty was assigned to the solar spectrum because it is extremely well established. The spectral fluxes in this article are from Wehrli (1985). There are more recent determinations and, in fact, the total solar constant has been revised downward from  Wehrli's 1367 to 1361 W/m^2 (Kopp and Lean, 2011). However, that is only a difference of 0.4% and the absolute value of flux does not change the albedo.

Section 5 discussed the constant factor of 6.97 applied to the spectroscopic data of McCord and Clark (1979) in order to match the MASCS data of Domingue et al. (2010). MASCS and MDIS are mutually consistent, and both are consistent with previous photometric observations of Mercury. A further validation of the MESSENGER data can be seen by comparing the MDIS geometric albedos from table 5 of Domingue et al. (2011) with the recently published values derived from Sloan system photometry of Mercury listed in table 7 of Mallama et al. (2017). Thus, the reflectance values of McCord and Clark are the data that require the application of a constant to bring them into conformance with other such data.

Three possible reasons for the discrepancy are: those authors defined reflectivity differently, there was an erroneous constant in one of their equations and the axis of their plot is mislabeled. Communications with the authors did not resolve the underlying cause of the discrepancy.

Turning now to geophysics, Mercury's effective blackbody temperature can be calculated from the spherical bolometric albedo and the solar flux at the planet's distance from the Sun. These quantities determine the amount of flux the planet absorbs per unit surface area as indicated by Eq. 2,

$$E = R_{sun} ( 1 - A_{sb} ) / 4$$

Eq. 2

where $E$ is the solar flux absorbed by Mercury, $R_{sun}$ is the solar flux incident upon the planetary disk, $A_{sb}$ is the spherical bolometric albedo and the factor $4$ is the ratio between the surface area of a globe and that of its projected disk.

Since absorbed and emitted flux are balanced, the effective blackbody temperature can be computed from the Stefan-Boltzmann law which relates emitted flux and temperature. The resulting value is 437 K. This is slightly lower than that listed by Williams (2016) which corresponds to his smaller value for the albedo.

Méndez and Rivera-Valentín (2017) recently developed a method for taking the eccentricity of a planet's orbit into account when computing its effective temperature. Méndez (private communication) indicates that in the case of Mercury with a spherical bolometric albedo of 0.088 and an orbital eccentricity of 0.206, the temperature is reduced from 437.5K to 436.3K.

Finally, it is worth noting that the planet's spectrum contrasts in its general form with those of other planets. For colder planets, the emitted and absorbed energy components can be neatly separated by observation because they fall in two different wavelength regimes. The absorbed energy depends on the bolometric albedo over the shorter wavelengths from the UV through the near-IR where there is significant solar flux; it is the factor '$R_{sun} ( 1 - A_{sb} )$' in Eq. 2. Meanwhile, the emitted energy is measured at much longer wavelengths in the thermal IR because the planets' surfaces are cool. For Mercury, on the other hand, the reflected and thermal spectra strongly overlap, due to its high day-side temperature, in the region from about one micron out to several microns.

8. Summary and conclusions

Three sources of published data on the reflectance and albedo of Mercury have been combined in order to determine the planet's spherical bolometric albedo. First is the already published albedo value of 0.081 derived from MESSENGER/MDIS data over the wavelength range from 0.43 to 1.01 μm by Domingue et al. (2011). Second is an array of MESSENGER/MASCS reflectance values reported by Domingue et al. (2010) which span wavelengths from 0.31 to 1.46 μm. Third is the set of reflectance values reported by McCord and Clark (1979) which cover from 0.65 to 2.50 μm. Additional extrapolation provided data spanning wavelengths from 0 to 4 μm.

Reflectance values from Domingue et al. (2010) and McCord and Clark (1979) were converted to fluxes. These, in turn, were compared to the solar spectral flux in order to determine the planet's weighted spherical albedo in five wavelength intervals. A sixth interval is that for which Domingue et al. (2011) determined a partial spherical bolometric albedo covering 59% of the solar flux. This analysis indicated that the wavelength-dependent albedo increases by a factor of ~10 from 0.019 in the interval 0.00 to 0.30 μm to 0.186 in the interval from 2.46 to 4.00 μm.

The reflected flux of Mercury from 0 to 4 μm was found to be 119.8 W/m^2, while the corresponding flux of the Sun is 1357.0. Dividing the flux of the planet by that of the Sun results in a spherical bolometric albedo of 0.088 which includes over 99% of the flux from the entire solar spectrum. The uncertainty of this albedo is estimated to be +/- 0.003. The corresponding effective blackbody temperature of Mercury is 437.5K. When the planet's orbital eccentricity is taken into account this is reduced to 436.3K.

Acknowledgments

D. L. Domingue commented on an earlier version of this paper and made many helpful suggestions. A. Méndez computed the effective blackbody temperature of Mercury that includes the effect of its elliptical orbit. Any errors in this paper, though, are solely the responsibility of the author.